# Ambient-Induced Selenium Segregation and Nanoparticle Formation in 2H-HfSe$_2$: An Experimental and Theoretical Study

Stefany P. Carvalho[1*], Guilherme S. L. Fabris[2*], Ana Carolina F. de Brito[1], Raphael B. de Oliveira[2,3*], Wesley Kardex C. de Oliveira[4], Catalina Ruano-Merchán[4], Carlos A. R. Costa[5], Luiz F. Zagonel[4], Douglas Galvão[2†], Ingrid D. Barcelos[1†].

[1]Brazilian Synchrotron Light Laboratory (LNLS), Brazilian Center for Research in Energy and Materials (CNPEM), Zip Code 13083-100, Campinas, São Paulo, Brazil.
[2]Applied Physics Department and Center for Computational Engineering & Sciences, State University of Campinas, Campinas, São Paulo 13083-970, Brazil
[3]Department of Materials Science and NanoEngineering, Rice University, Houston, TX 77005, USA.
[4]Gleb Wataghin Physics Institute, University of Campinas (Unicamp), 13083-859 Campinas, São Paulo, Brazil
[5]Brazilian National Nanotechnology Laboratory (LNNano), Brazilian Center for Research in Energy and Materials (CNPEM), Zip Code 13083-100, Campinas, São Paulo, Brazil.

*These authors contribute equally
†Corresponding-Author: ingrid.barcelos@lnls.br; galvao@ifi.unicamp.br;

**Abstract**

We investigate the air-induced degradation of few-layer hafnium diselenide (HfSe$_2$) through combined experimental and theoretical approaches. AFM and SEM reveal the formation of selenium-rich spherical features upon ambient exposure, while EDS confirms Se segregation. Ab initio molecular dynamics simulations show that Se atoms migrate to flake edges and that O/O$_2$ exposure leads to selective Hf oxidation, breaking Se–Hf bonds and expelling Se atoms. No stable Se–O bonds are observed, indicating structural reorganization rather than oxidation. These findings emphasize the material's instability in air and the importance of encapsulation for preserving HfSe$_2$ in practical applications. Scanning tunneling spectroscopy confirms the semiconducting character of the nanoparticles, with an electronic bandgap compatible with that of elemental Se. These results highlight the critical role of lattice defects and oxidation dynamics in the degradation process and underscore the need for encapsulation strategies to preserve the integrity of HfSe$_2$-based devices.

**Introduction**

Research on two-dimensional (2D) materials has grown exponentially due to their remarkable electronic, optical, and mechanical properties, as well as the possibility of controlled stacking to form van der Waals heterostructures with customizable functionalities [1], [2], [3]. Among these, transition metal dichalcogenides (TMDs), with the general formula MX$_2$ — where M is a transition metal (e.g., Mo, W, Hf) and X is a chalcogen (e.g., S, Se, Te) — stand out as a diverse class of layered semiconductors featuring tunable band gaps, high carrier mobilities, and strong interlayer coupling [4], [5], [6], [7], [8]. These characteristics make TMDs attractive for various applications, including field-effect transistors (FETs), sensors, energy storage, and flexible optoelectronic devices [9], [10], [11]. While considerable attention has been given to disulfides and diselenides such as MoS$_2$, WS$_2$, MoSe$_2$, and WSe$_2$, other compounds remain comparatively underexplored.

Hafnium diselenide (HfSe$_2$), in particular, has emerged as a promising candidate for FETs due to its moderate indirect electronic band gap (~1 eV), comparable to that of silicon, and the presence of a native high-κ oxide (HfO$_2$), which is advantageous for gate engineering [12], [13], [14]. However, the practical deployment of HfSe$_2$-based devices hinges on a detailed understanding of its environmental stability, as exposure to ambient conditions can lead to chemical reactions, defect generation, and morphological changes that degrade device performance [15], [16], [17]. Although $HfSe_2$-based devices have demonstrated high on/off ratios and fast optoelectronic responses, the experimentally reported carrier mobilities remain significantly lower than theoretical predictions. This discrepancy has been attributed to a high density of intrinsic defects and pronounced sensitivity to environmental exposure, which together compromise the structural and electronic integrity of the material [17]. Recent studies indicate that $HfSe_2$ undergoes rapid degradation in air, potentially involving hafnium oxidation and/or selenium segregation [12], [17]. However, the microscopic mechanisms underlying these transformations—particularly the nucleation and evolution of Se-rich surface features—remain poorly understood.

In this work, we present a comprehensive experimental and theoretical investigation of the air-induced degradation of few-layer $HfSe_2$. Using atomic force microscopy (AFM), scanning electron microscopy (SEM), and energy-dispersive X-ray spectroscopy (EDS), we observe the formation of selenium-rich spherical structures on the flake surface. Scanning tunneling microscopy and spectroscopy (STM/STS) further reveal the semiconducting character of these nanoparticles, with an electronic bandgap consistent with elemental selenium, suggesting a segregation-driven mechanism and surface reorganization rather than simple oxidation. To elucidate the atomistic origin of this process, we employ *ab initio* molecular dynamics simulations to uncover the driving forces behind droplet formation and their preferential nucleation at flake edges. Our results integrate experimental evidence with theoretical modeling, providing new insights into the degradation pathways of reactive 2D materials and into forming strategies for environmental stabilization.

**Materials and Methods**

**Sample preparation**

Few-layer $HfSe_2$ samples were prepared by mechanical exfoliation from a bulk crystal synthesized and supplied by 2D Semiconductors. The exfoliation was performed using low-adhesion silicone adhesive tape to separate the layers held together by van der Waals forces. The exfoliated flakes were then transferred onto $Si/SiO_2$ substrates by pressing the tape against the surface. Optical microscopy was employed for preliminary identification of suitable flakes, based on optical contrast. This method relies on interference and absorption effects arising from the flake thickness and the underlying substrate, allowing estimation of the number of layers through visible color variations.

**Morphological and Chemical Characterization**

Atomic Force Microscopy (AFM) measurements were conducted using a NX10 scanning probe microscope (Park Systems) operating in intermittent contact mode. The system was equipped with PPP-FMR probes (Nanosensors). All measurements were performed under controlled environmental conditions (relative humidity up to 55%, and temperature maintained around 25 °C) within an acrylic enclosure. This setup enabled non-destructive, high-resolution imaging of surface topography, roughness, and potential contamination.

Scanning Electron Microscopy (SEM) analyses were performed using a Thermo Fisher Scientific Helios 5 PFIB CXe DualBeam system, operated in high-vacuum mode at 15–20 kV, 0.1 nA, and a magnification of 60,000x. The samples already exhibited spherical features prior to imaging. Elemental composition was assessed using Energy-Dispersive X-ray Spectroscopy (EDS) with an Oxford Instruments X-Max detector, enabling the mapping of chemical variations across the surface as the beam scanned the sphericity regions.

**Scanning Tunneling Microscopy and Spectroscopy (STM/STS)**

STM and STS measurements were performed under UHV conditions at room temperature (RT) using a modified RHK PanScan FlowCryo microscope [18]. With this setup, imaging can be associated with electronic and optical spectroscopies. Prior to STM and STS experiments, the sample was annealed at 383 K for 12 hours under a vacuum of $10^{-7}$ Torr for cleaning atmospheric contaminations. The background pressure of the experimental chamber was about $1 \times 10^{-10}$ Torr. STM images were performed in constant current mode using a tungsten tip prepared by electrochemical etching with NaOH solution and processed with WSxM software [19]. For STS measurements, 200 spectra were acquired at room temperature after tip stabilization at −1 V and 2 nA, with the feedback loop subsequently disabled.

**Computational Methods**

We have carried out Density Functional Theory (DFT) [20]-based simulations using the SIESTA code [21], [22]. We have employed the Perdew-Burke-Ernzerhof (PBE) [23] exchange-correlation functional, which uses the Generalized Gradient Approximation (GGA) combined with a double-polarized (DZP) basis set, which is composed of numerical orbitals. In all simulations, a mesh cutoff of 400 Ry and a Γ-centered Monkhorst−Pack grid [24] composed of 2x2x1 k-points. To build our model, we created two $HfSe_2$ nanostructures: the first one was a $HfSe_2$ nanoribbon (periodic only along the y-direction) with a unit cell composed of 70 atoms (*a* = 35.15 Å, *b* = 22.84 Å, *c* = 30.00 Å, α = β = 90.0° and γ = 120.13°) and a monolayer (periodic in x- and y-direction) which has a unit cell of 27 atoms (a = 11.37 Å, b = 11.42 Å, c = 30.00 Å, α = β = 90.0° and γ = 120.13°). For both nanostructures, a buffer vacuum region of 30 Å along the z-axis was used to prevent spurious interactions between the structure and its periodic (mirror) images. Both structures were optimized (at 0K) until their residual forces were smaller than 0.05 eV/Ang, and during each self-consistency iteration, we considered that the convergence was achieved when the difference between the input and output of each element of the density matrix was lower than $10^{-3}$. To evaluate their dynamical structural properties and investigate the oxygen interaction with the monolayer, we have also carried out *ab initio* DFT molecular dynamics (AIMD) simulations. Our simulations were performed using an NVT ensemble. The AIMD runs were performed for approximately three ps with a one-fs time step, during which the temperature was maintained at 300 K through a Nosé-Hoover thermostat [21], [22]. For convergence, we use the same parameters as those used during geometrical DFT optimizations.

**Results and Discussion**

**Figure 1** provides an overview of the structural and morphological characteristics of hafnium diselenide ($HfSe_2$). The crystal used in this study, acquired from 2D Semiconductors (**Figure 1a**), is in the 2H phase, which is the thermodynamically most stable form of many transition-metal dichalcogenides (TMDs), including $MoS_2$, $WS_2$, and $HfSe_2$ [25], [26]. This phase consists of a layer of hafnium (Hf) atoms sandwiched between two layers of selenium (Se) atoms, as illustrated in the side and top views of the unit cell in **Figure 1b**, which shows the typical hexagonal arrangement of this two-dimensional material. The samples were prepared by mechanical exfoliation of the fresh crystal in a controlled environment with low humidity and reduced oxygen levels, to avoid undesired surface reactions. Optical microscopy images of freshly exfoliated flakes (**Figure 1c**) reveal few-layer $HfSe_2$ nanosheets with uniform morphology and smooth surfaces. However, after one month of ambient exposure, the same flake exhibits dark spherical features across its surface (**Figure 1d**), indicating a degradation process triggered by environmental conditions.

This behavior is consistent with previous studies [12], [27], [28], [29],[35], which report oxidation and morphological instability in TMDs exposed to air. Differences in degradation times reported in the literature may arise from variations in environmental parameters such as humidity and temperature, as in the case of Yao et al. (2018) [12], who reported morphological changes and degradation of TMDs, including $HfSe_2$, in open atmospheres. The differences in degradation times observed may be attributed to variations in experimental conditions, such as humidity and temperature. To further investigate the environmental stability of $HfSe_2$, a systematic study was conducted using atomic force microscopy (AFM), scanning electron microscopy (SEM), and energy-dispersive X-ray spectroscopy (EDS). These techniques enabled a comprehensive analysis of surface morphology and chemical composition throughout the aging process, allowing correlation between structural evolution and ambient exposure.

Initial AFM topographic images (**Figure 1e**) show a smooth and homogeneous surface, as expected for freshly exfoliated samples stored under vacuum and at relative humidity below 5%. To simulate the natural aging process, the samples were progressively exposed to ambient humidity levels exceeding 55%. AFM topographic images were performed daily over a one month period to monitor the topographical evolution in high resolution. **Figures 1f**-**1i** display AFM topography images of the flake surfaces, showing the gradual formation and growth of spherical features after 8 hours, 1 day, 2 days, and 1 week of exposure, respectively. After one month (**Figure 1j**), the features reach diameters of approximately 90 nm. **Figure k** shows an exponential growth curve of the average sphericity diameter, as a function of exposure time, highlighting the accelerated kinetics of this surface transformation under ambient humidity. These results suggest that humidity plays a critical role in the formation and expansion of these features, potentially mediated by structural defects such as selenium vacancies.

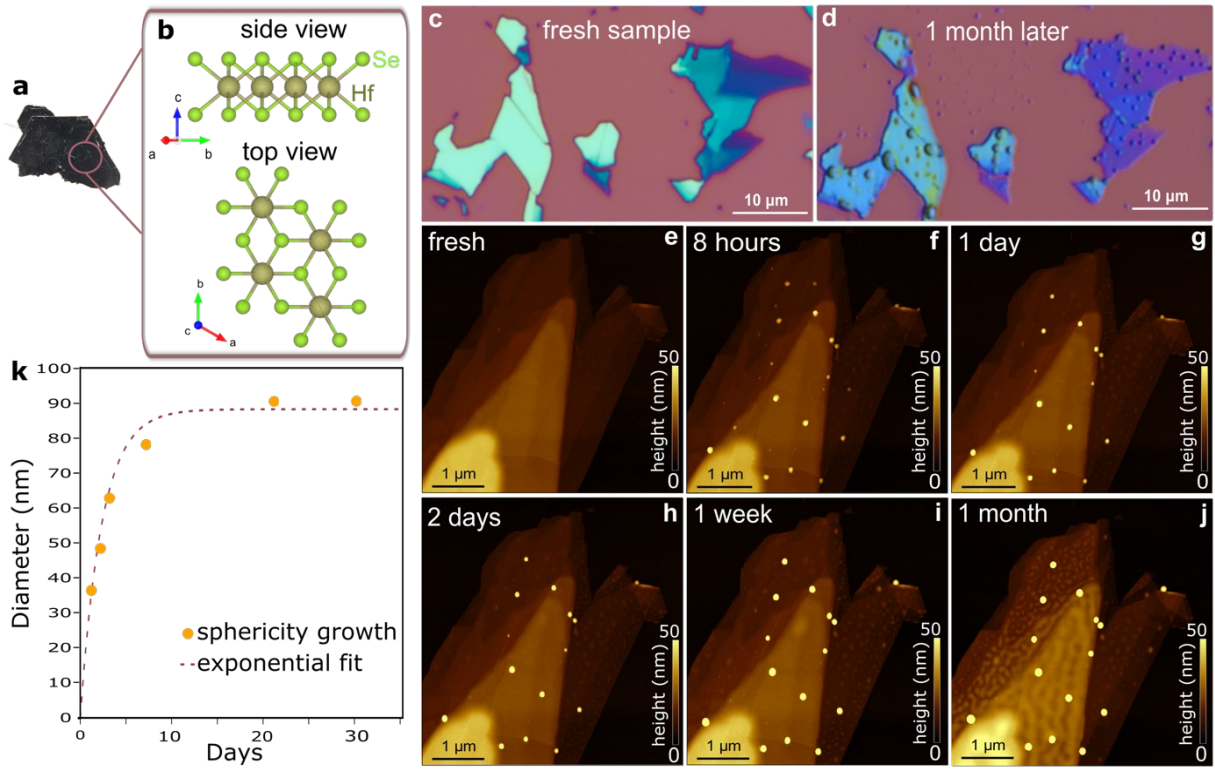

**Figure 1. Structural, morphological, and time-dependent evolution of few-layer HfSe₂ under ambient conditions.** (a) Optical image of the HfSe₂ crystal acquired from 2D Semiconductors. (b) Schematic representation of the 2H phase of HfSe₂, showing a side view with the Se–Hf–Se layered structure and a top view highlighting the hexagonal arrangement. (c) Optical microscopy image of a freshly exfoliated HfSe₂ flakes deposited on SiO₂/Si substrate. (d) Optical image of the same flakes after one month of air exposure, showing the emergence of dark spherical features across the surface. (e–j) Atomic Force Microscopy (AFM) topographic images of the flake surface at increasing time intervals: (e) fresh, (f) 8 hours, (g) 1 day, (h) 2 days, (i) 1 week, and (j) 1 month. These images show the progressive formation and growth of spherical structures on the surface, with diameters reaching ~90 nm after one month. (k) Exponential fit of the average sphere diameter as a function of exposure time, indicating growth saturation.

The sphericities formed on the surface of HfSe₂ flakes were investigated in more detail by Scanning Electron Microscopy (SEM) and Energy Dispersive X-ray Spectroscopy (EDS), as shown in **Figure 2**. The SEM images (**Figures 2a-b**) revealed that these structures initially appear at sample fractures, areas more susceptible to chemical reactions due to increased surface energy and greater exposure to the environment [30], [31]. Over time, these sphericities spread across the entire surface of the flake, corroborating previously made degradation observations. The chemical analysis by EDS, shown in the elemental mapping (**Figure 2b**), indicated a higher concentration of selenium in the spherical regions, compared to other elements present, such as oxygen and hafnium, which are mainly distributed in the surrounding matrix. The line profile analysis (**Figure 2c**) reinforced this observation, showing a significant increase in selenium in the spheres compared to the rest of the sample.

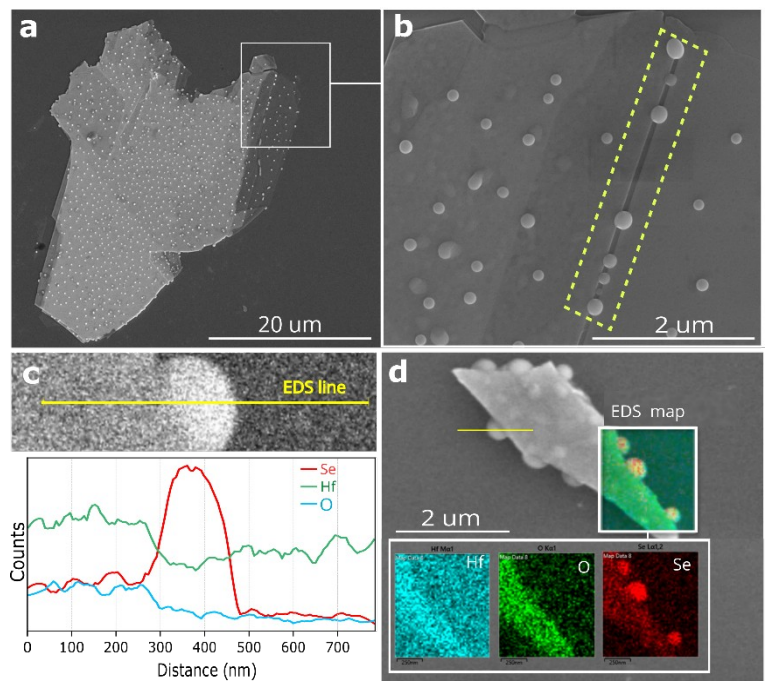

**Figure 2. SEM and EDS analysis of HfSe$_2$ degradation and selenium nanoparticle formation.** (a) Low-magnification SEM image showing the overall morphology of a HfSe$_2$ flake after one month of air exposure, with widespread formation of spherical features across the surface. (b) High-magnification SEM image highlighting the alignment of spherical structures along a fracture line (yellow dashed box), suggesting preferential nucleation at high-energy regions. (c) EDS line profile acquired along the yellow trajectory shown in the upper panel, revealing a strong selenium signal localized at the center of the spheres, in contrast with the surrounding matrix enriched in hafnium and oxygen. (d) EDS elemental mapping of a selected flake region, with the green box highlighting localized selenium enrichment at the edges of a crack. The elemental maps (bottom panels) show the spatial distributions of Hf, O, and Se, confirming selenium accumulation within the spherical features.

These results suggest a possible selective oxidation, as indicated in previous studies, in which HfSe$_2$ undergoes oxidation under environmental conditions, forming selenium-enriched byproducts [12], [13], [35]. However, the absence of significant content within the spheres contradicts the hypothesis of direct oxidation. This finding supports a degradation mechanism dominated by selenium segregation and chemical/structural reorganization, rather than oxide formation. The high selenium concentration within the spherical features, along with its depletion in the surrounding matrix, may point to the formation of selenium nanoparticles, possibly triggered by disproportionation reactions or selective selenium volatilization. Thus, the observed phenomenon reflects not merely oxidation, but rather a redistribution of selenium atoms facilitated by point defects such as selenium vacancies. A likely explanation for this behavior is the partial dissociation of the $HfSe_2$ lattice, promoting the detachment of selenium atoms that subsequently agglomerate on the surface. Selenium vacancies—formed when Se atoms are removed from the crystal—destabilize the structure and create reactive sites that enhance interactions with environmental species such as oxygen and water vapor, thereby accelerating the morphological transformation.

To elucidate the mechanisms behind this degradation process, we performed AIMD simulations. **Figures 3a-b** show the initial and final configuration, where four additional selenium atoms (in gold) were deposited on top of the nanoribbon —three near the edges and one near the center. In a first step, to investigate the preferential nucleation of selenium-rich features at flake edges, we deposited Se atoms in both central and edge regions of a $HfSe_2$ nanoribbon. The simulations revealed that free selenium atoms spontaneously migrate toward the edges, where the local potential energy is minimized. This result reinforces the hypothesis that flake edges act as energetically favorable sites for the nucleation and growth of selenium nanoparticles. The additional four selenium atoms are colored differently from those present in the $HfSe_2$ nanoribbon for better visualization; moreover, a rectangular blue box has been drawn to illustrate the unit supercell. After we deposited the Se atoms at different locations, we performed the AIMD simulation for 3000 fs. The final configuration of the system is shown in **Figure 3b**. Regarding the Se atoms closer to the edge, due to the absence of periodicity along the x-axis, those Se start to move to the position that they would occupy if there had hafnium atoms in that region, so that they maintain the structural integrity present in the HfSe2 monolayer, indicating a possible explanation for the creation of nanoparticles at the borders, as the free selenium atoms tends to diffuse to those regions with smaller potential energy barriers, i. e. , the borders. Also, the Se in the middle of the nanoribbon moves through the surface toward the edges; this diffusion reinforces the evidence that the selenium atoms, when free in a finite $HfSe_2$ structure, easily diffuse to the borders. See videos in the Supplementary Information for a better visualization of the whole process.

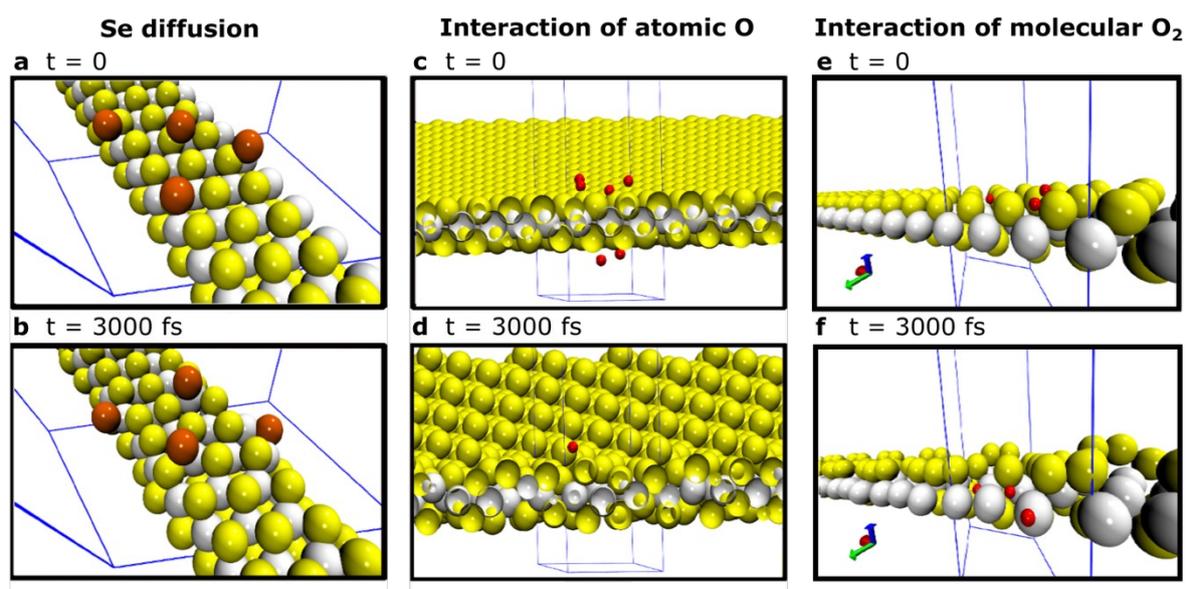

**Figure 3. AIMD simulations illustrating key atomistic mechanisms underlying the air-induced degradation of $HfSe_2$.** (a), (c) and (e) Initial configuration of the simulation for an $HfSe_2$ nanoribbon. (b), (d) and (f) Final configuration after 3000 fs of simulation at 300 K. (a-b) Se diffusion on a $HfSe_2$ nanoribbon: the free Se atoms preferentially migrate toward the edges, where potential energy barriers are lower. (c-d) Interaction of atomic O with a $HfSe_2$ monolayer: oxygen atoms diffuse into the lattice, forming Hf–O bonds and destabilizing the structure via Se–Hf bond breaking and Se expulsion. (e-f) Interaction of molecular $O_2$ with a $HfSe_2$ monolayer: $O_2$ dissociates upon adsorption, and the resulting O atoms bind to Hf, inducing further Se displacement and structural rearrangements.

To further explore the role of oxidation, we modeled the interaction of atomic oxygen with a HfSe$_2$ monolayer. As shown in **Figures 3c-3d**, there are some positions occupied by the oxygen atoms, which allowed them to interact and react with the Hf atoms present in the "middle" of the two Se layers. In other words, the O atoms can diffuse into the monolayer due to the high affinity of Hf for oxygen, causing their oxidation. Furthermore, during the Hf oxidation, the Se-Hf bonds are broken, and the Se atoms are 'expelled' from the structure, and they can diffuse over the monolayer toward its borders. Therefore, due to the oxidation of Hf in the system, the available oxygen does not tend to form selenium oxide, but Hf oxides. This can be clearly seen towards the end of the simulation, where even using eight oxygen atoms, only one selenium-oxygen bond was formed, consistent with the formation of Se-rich surface features observed experimentally. See videos in the Supplementary Information for a better visualization of the whole process.

Additionally, we have carried out a third type of simulation based on the fact that the region between two consecutive selenium atoms is an area capable of capturing O. Therefore, we have placed two O$_2$ molecules in two distinct regions of the monolayer. As shown in **Figure 3e**, this configuration enabled O$_2$ to access reactive regions without bias from structural edges. After 3000 fs of AIMD simulation at room temperature, we observed the dissociation of both O$_2$ molecules into atomic oxygen. As depicted in **Figure 3f**, the resulting O atoms preferentially bonded to hafnium atoms, triggering the breakage of Hf–Se bonds and again expelling Se atoms from the lattice. This behavior is consistent with the previously observed oxidation mechanism, further confirming that oxygen exposure leads to selective Hf oxidation and subsequent structural destabilization. No stable Se–O bonds were formed, once again indicating that Se atoms are not oxidized, but rather displaced and segregated due to lattice disruption.

To further investigate the energetics of the oxidation process, we performed a comparative AIMD simulation analogous to the previous one, but with two $O_2$ molecules initially placed at 15 Å above the $HfSe_2$ monolayer to mimic a more physically realistic adsorption scenario. During the simulation, we tracked the system's total energy evolution to assess the thermodynamic favorability of the oxidation process. As shown in **Figure** 4, the interaction proceeds through a rapid physisorption phase, followed by dissociation of the $O_2$ molecules and formation of Hf–O bonds. This leads to Se atom displacement and the appearance of surface rearrangements. The associated energy profile reveals an initial decrease due to adsorption, a slight increase during bond reorganization, and subsequent stabilization — indicating that oxidation is energetically favorable and dynamically accessible at room temperature. Representative snapshots along the simulation timeline illustrate the structural transformations that accompany this process. Following a 3000 fs MD simulation, we computed the total energy difference between the non-interacting (isolated) and interacting systems, which are shown in **Figure 4**. This energy variation provides insight into the thermodynamic favorability of $O_2$ adsorption and the subsequent oxidation process at the $HfSe_2$ surface.

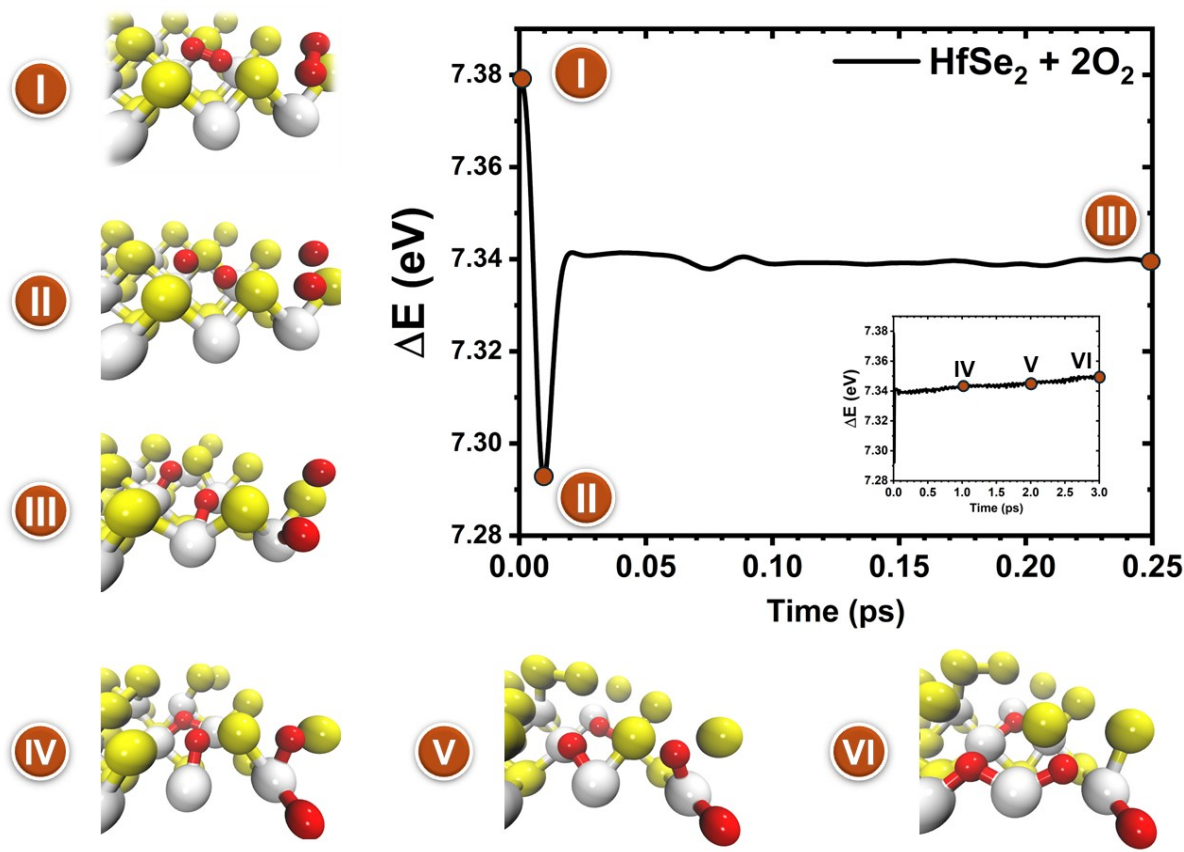

**Figure 4: AIMD simulation revealing the energetic evolution of the HfSe₂ monolayer during oxidation by O₂ molecules.** Energy variation profile comparing HfSe₂ in the presence of two O₂ molecules versus isolated HfSe₂. Representative AIMD snapshots illustrate key stages of the simulation timeline. Hafnium, selenium, and oxygen atoms are represented by white, yellow, and red spheres, respectively.

From **Figure 4**, it can be observed that once the O₂ molecules begin interacting with the HfSe₂ surface (stage I), there is an initial decrease in total energy due to physorption. This is followed by a gradual energy increase as the O–O bonds break (stage II) and start interaction with Se and Hf atoms, marking the onset of surface reactivity, leading to Hf oxidation and 'expelled' Se (stage IV). After completing this stage, the system remains at equilibrium (stages V and VI). As the system approaches energetic stabilization, indications of Se atom clustering begin to emerge around the reaction sites, suggesting a rearrangement of the surface composition driven by the oxidation dynamics. See videos in the Supplementary Information for a better visualization of the whole process.

**STM/STS Measurements on Selenium Nanostructures**

The electronic structure of the HfSe$_2$ nanoparticles was investigated by scanning tunneling microscopy (STM) and scanning tunneling spectroscopy (STS). **Figure 5a** shows the STM topography of a representative nanoparticle formed at the HFse$_2$ surface after air exposure, where the STM tip was positioned for the acquisition of local tunneling spectra. A total of 200 I–V curves (orange) and their corresponding dI/dV spectra (black) were acquired with the feedback loop disabled after tip stabilization at a setpoint of 2 nA and −1 V, as presented in **Figure 5b**. The onset of the differential conductance at negative and positive sample biases is assigned to the valence band maximum (VBM) and conduction band minimum (CBM), respectively. Linear extrapolation of these edges yields an apparent electronic band gap ($E_g$) of ≈ 1.7 eV. This value is inconsistent with the wide-gap insulating character of HfO$_2$, where STS measurements on nanoislands reported $E_g$ = (5.78 ± 0.13) eV [32]. Also, our spectra show an electronic gap much higher than expected for HfSe$_2$. Yue *et al.* reported an STS-derived band gap of ~1.1 eV for HfSe$_2$, with the Fermi level near the conduction band due to selenium vacancies [33]. Likewise, STS mapping revealed gaps of ~1.0 eV on pristine surfaces HfSe$_2$ and up to ~1.25 eV at defect sites, reflecting spatial LDOS fluctuations induced by native defects [12].

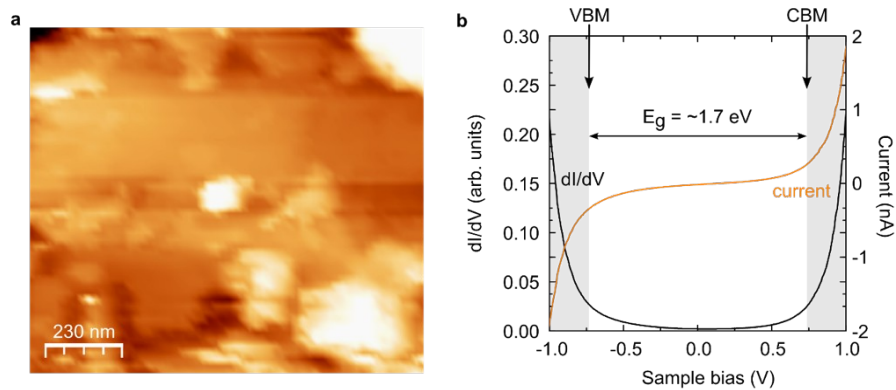

**Figure 5: STM and STS measurements of nanoparticles on a HfSe₂ flake.** (a) STM topography of the HfSe₂ surface showing an individual nanoparticle, likely corresponding to segregated selenium (tunneling parameters: $V_{bias}$ = 2 V, $I$ = 0.5 nA). (b) Corresponding STS (I–V and dI/dV) spectra acquired with the STM tip positioned on the nanoparticle. The feedback loop was disabled after tip stabilization at 2 nA and −1 V. The onsets of the differential conductance at negative and positive sample biases define the valence band maximum (VBM) and conduction band minimum (CBM), yielding an apparent electronic band gap of ~1.7 eV. This value is incompatible with HfO₂ but consistent with selenium-rich nanoparticles.

By contrast, our measured gap is compatible with the electronic band gap of trigonal selenium (~1.5 eV [34]), suggesting that the nanoparticle electronic response is that of Se instead of HfSe₂ or HfO₂. This observation is further evidence that the observed nanoparticles are not oxide phases but rather Se segregation formed at the surface. Furthermore, the finite *dI/dV* intensity observed within the nominal gap region indicates the presence of intragap states, likely associated with structural defects, Se vacancies, or unintentional contaminants at the nanoparticle surface. Such states would naturally account for the deviation from the ideal Se band structure and the variability in the extracted gap values.

**Conclusions**

In this work, we presented a comprehensive experimental and theoretical investigation of the air-induced degradation mechanisms in few-layer HfSe₂. Morphological analyses by AFM and SEM revealed the progressive formation of selenium-rich spherical features on the flake surface upon ambient exposure.

EDS mapping confirmed a strong Se enrichment in these regions, while STM/STS measurements revealed an electronic bandgap consistent with semiconducting selenium phases. The presence of intragap states in the tunneling spectra suggests that these nanoparticles are strongly influenced by structural defects and/or surface contamination, supporting their identification as segregated Se nanostructures rather than oxide-derived islands. These findings indicate that the degradation mechanism is driven by selenium segregation and surface reconstruction, rather than simple oxidation. To rationalize the experimental results, we performed ab initio molecular dynamics simulations that revealed the spontaneous diffusion of free selenium atoms toward the flake edges, as well as the preferential oxidation of hafnium upon O or $O_2$ exposure. The simulations showed that oxygen atoms break Hf–Se bonds, triggering selenium displacement without the formation of stable Se–O bonds. The associated energy profiles confirmed that these processes are thermodynamically favorable and dynamically accessible at room temperature, leading to structural rearrangements and the emergence of Se-rich nanostructures. Together, these results demonstrate that although $HfSe_2$ is easily exfoliated and possesses promising properties for integration into van der Waals heterostructures, its surface is highly reactive under ambient conditions. Consequently, strategies such as encapsulation or inert-atmosphere storage are essential to preserve its structural and electronic integrity in practical applications.

**Author contributions**

**Conceptualization:** Ingrid D. Barcelos and []. **Formal analysis:** Stefany P. Carvalho, Carlos A. R. Costa, []. **Investigation:** Stefany P. Carvalho, [] **Simulations:** Raphael B. de Oliveira, Guilherme S. L. Fabris and Douglas Galvão. **Project administration:** Ingrid D. Barcelos. **Resources:** []. **Writing – original draft:** Ingrid D. Barcelos and Stefany P. Carvalho **Writing – review & editing:** Ingrid D. Barcelos, Stefany P. Carvalho, Ana Carolina F. de Brito, Raphael B. de Oliveira, Guilherme S. L. Fabris and Douglas Galvão, []

**Conflicts of interest**

There are no conflicts to declare.

**Data availability**

All data that support the findings of this study are included within the article (and any supplementary files).

**Acknowledgements**

The authors thank all Brazilians and also thank the Brazilian Synchrotron Light Laboratory (LNLS) and the Brazilian Nanotechnology National Laboratory (LNNano), both part of the Brazilian Center for Research in Energy and Materials (CNPEM), a private non-profit organization under the supervision of the Brazilian Ministry for Science, Technology, and Innovations (MCTI), for the preparation and characterization of the samples — LAM-PFIB (Proposal No. 20252386), LAM-L2D (Proposal No. 20252214) and LNNano (Proposal No. 20241094). IDB and ACFB acknowledge financial support from CNPq (grant number 351538/2023 − 2). Guilherme S. L. Fabris acknowledges the São Paulo Research Foundation (FAPESP) fellowship (process number 2024/03413-9). Raphael B. de Oliveira thanks the National Council for Scientific and Technological Development (CNPq) (process numbers 151043/2024-8 and 200257/2025-0), Douglas S. Galvão acknowledges the Center for Computing in Engineering and Sciences at Unicamp for financial support through the FAPESP/CEPID Grant (process number 2013/08293-7). We thank the Coaraci Supercomputer Center for computer time (process number 2019/17874-0).